\documentclass[conference]{IEEEtran}

\IEEEoverridecommandlockouts
\usepackage{cite}
\usepackage{amsmath,amssymb,amsfonts}
\usepackage{algorithmic}
\usepackage{graphicx}
\usepackage{float}
\usepackage[export]{adjustbox}
\usepackage{textcomp}
\usepackage{xcolor}
\usepackage{multicol}
\usepackage{multirow}

\newcommand\T{\rule{0pt}{2.6ex}}       
\newcommand\B{\rule[-1.5ex]{0pt}{0pt}} 

\def\BibTeX{{\rm B\kern-.05em{\sc i\kern-.025em b}\kern-.08em
    T\kern-.1667em\lower.7ex\hbox{E}\kern-.125emX}}
\begin{document}

\clearpage
\onecolumn 
\pagestyle{empty} 
{© 2022 IEEE.  Personal use of this material is permitted.  Permission from IEEE must be obtained for all other uses, in any current or future media, including reprinting/republishing this material for advertising or promotional purposes, creating new collective works, for resale or redistribution to servers or lists, or reuse of any copyrighted component of this work in other works.
DOI: 10.1109/ISQED54688.2022.9806292

\hfill
\hfill
\hfill

Please cite this paper as following:

\hfill
\hfill
\hfill

M. Eslami, T. Ghasempouri and S. Pagliarini, "Reusing Verification Assertions as Security Checkers for Hardware Trojan Detection," 2022 23rd International Symposium on Quality Electronic Design (ISQED), Santa Clara, CA, USA, 2022, pp. 1-6, doi: 10.1109/ISQED54688.2022.9806292.}
\twocolumn 

\title{Reusing Verification Assertions as Security Checkers for Hardware Trojan Detection}
\author{\IEEEauthorblockN{ Mohammad Eslami, Tara Ghasempouri and Samuel Pagliarini }
\IEEEauthorblockA{\textit{Department of Computer Systems} \\
\textit{Tallinn University of Technology (TalTech)}\\
Tallinn, Estonia \\
Email: \{FirstName.LastName@taltech.ee\}} }
 \vspace{-25mm}

\maketitle

\begin{abstract} 
 Globalization in the semiconductor industry enables fabless design houses to reduce their costs, save time, and make use of newer technologies. However, the offshoring of Integrated Circuit (IC) fabrication has negative sides, including threats such as Hardware Trojans (HTs) -- a type of malicious logic that is not trivial to detect. One aspect of IC design that is not affected by globalization is the need for thorough verification. Verification engineers devise complex assets to make sure designs are bug-free, including assertions. This knowledge is typically not reused once verification is over. The premise of this paper is that verification assets that already exist can be turned into effective security checkers for HT detection. For this purpose, we show how assertions can be used as online monitors. To this end, we propose a security metric and an assertion selection flow that leverages Cadence JasperGold Security Path Verification (SPV). The experimental results show that our approach scales for industry-size circuits by analyzing more than 100 assertions for different Intellectual Properties (IPs) of the OpenTitan System-on-Chip (SoC). Moreover, our detection solution is pragmatic since it does not rely on the HT activation mechanism. 
\end{abstract}


\section{Introduction and Background}\label{a}

Over the last decades, the Integrated Circuit (IC) industry has experienced significant changes in the fabrication process due to globalization. This globalization has made the companies define new strategies to reduce their costs in the IC supply chain. Hence, today, it is very unlikely for a design house to fabricate a circuit. Instead, the fabrication process is outsourced to third-party vendors. This opens an opportunity for an attacker to replace some parts of the original circuit with altered ones, or even to remove some logic in the design process \cite{c10}. This threat is generally referred to as Hardware Trojans (HTs). Hardware Trojans are known as one of the greatest threats against the trustworthiness of ICs, and they have raised serious concerns about the reliability and security of digital designs \cite{c3}. If an IC is utilized in a product/system while Trojan(s) remain there, it may lead to functionality change, reliability degradation/denial of service, and information or data leakage \cite{tiago}.

Unfortunately, typical test and verification tasks are not sufficient to detect HTs that are inserted during the fabrication stage \cite{c10} (i.e.,  fabrication-time attacks). Many accepted approaches exist to enable HT detection \cite{dis, b11, b12, c3, c, b}. In these approaches, it is tried to detect the HTs either by disruptive methods or non-disruptive ones. In disruptive methods, the IC is being de-masked and if necessary, de-layered. Then it goes through the investigation using electron microscopes and special measurement equipment to check if it is the same as designed \cite{dis}. Non-disruptive methods focus on catching the unwanted behavior of the circuit using analytical or formal methods. Mostly used techniques in non-disruptive methods include side-channel analysis and online monitoring. Side-channel analysis techniques are based on the concept of checking the different parametric characteristics of the circuit such as power and timing and looking for a deviation from the expected signatures to detect Trojans \cite{b11, b12}. The major drawback of the side-channel analysis technique is that the obtained results from the analysis can be confused with the process variation. Some approaches try to enhance the detection probability of this technique by applying Automatic Test Pattern Generation (ATPG) or dummy flip-flop insertion \cite{c3, c}.

On the other hand, online monitoring techniques rely on embedding checker circuits in different locations of the design to catch the unwanted behavior of the system \cite{b, s, t}. One of the popular approaches for building these checker circuits is to use assertions \cite{asr}. Assertions precisely describe the expected behavior of the circuit and help to check if there is a deviation between the intent and the actual behavior of the design using simulation or formal methods \cite{asr}. Assertion-Based Verification (ABV) techniques are mostly implemented by writing assertion checkers in Property Specification Language (PSL) \cite{a6} or SystemVerilog \cite{a8}. 

Although online monitoring techniques offer a high detection coverage, they impose significant overheads on the circuit. In recent years, it is tried to decrease the overheads while keeping the detection coverage at the maximum level, but still, the trade-off between the detection coverage and the imposed overheads is unfavorable \cite{t}. 

Since most of the time and efforts spent on the design and verification processes do not contribute to achieving HT detection, a considerable amount of design knowledge (i.e. test benches, coverage, metrics, assertions) is generated and then not re-utilized. In this paper, we propose a methodology for selecting the assertions that have already been written by verification engineers (for functional verification) and explain how to reuse them for security purposes (i.e., HT detection). Reusing the available data seems to be a wiser choice rather than spending similar (or more) time and resources on complex detection schemes. For this purpose, we present a new metric called \emph{security coverage} to evaluate the efficiency of online checkers in detecting Trojans considering the amount of overhead imposing on the circuit. This metric helps to remove assertions that are not helpful for HT detection from the circuit in an automated fashion while there is no need for detailed knowledge of the circuit.

The remainder of the paper is organized as follows: In Section \ref{b}, we explain the prerequisites for converting assertions to security checkers and introduce a new security coverage metric. The effectiveness of the selected checkers is studied in Section \ref{c}. Section \ref{d} explains a methodology to optimize the security checker list, and experimental results are presented in Section \ref{e}. Finally, Section \ref{f} concludes the paper.

\section{Assertions as Trojan Detectors}\label{b}

In this section, we answer the question \emph{can assertions be devised for detecting Trojans?} For this purpose, we study the B19-T500 benchmark from Trust-Hub \cite{tr1} which is a Trojan-inserted version of the B19 circuit from the ITC’99 benchmark suite \cite{itc}. Trust-Hub benchmarks provide an opportunity for developers to verify the effectiveness of their HT detection schemes since the Trojans can be considered as a representative for real ones due to their small sizes and rare triggering conditions \cite{tr3}. This implies the HTs remain hidden during standard verification checks \cite{tr2}.

More precisely, the B19 benchmark consists of four copies of the Viper processor, and the Trojan circuit is located inside each Viper processor. The Trojan is triggered by a counter which counts a specific vector and resets with another specific vector. If the counter gets a value between 3'b100 and 3'b110, the Trojan is triggered. The Trojan payload modifies the bits of the Instruction Register (IR) of the embedded Viper processor and changes the functionality of the circuit \cite{tr1}. 

\subsection{Prerequisite of good Security Checker}

The ideal conditions for assertions to be considered as good security checkers are: 1) they should not impose significant overheads on the circuit once they are synthesized (many of them may be needed in complex designs to reach the high detection coverage), and 2) they should not be limited in scope. Assertions that only check some local signals (i.e., checking if one of the specific bits of a register is 0 or 1) are rarely interesting. Instead, assertions that capture a high-level behavior are preferred. From now on, we call these assertions ``top-level assertions”. 

To clarify this concept, we have defined a set of manually written assertions that satisfy the aforementioned conditions. While the easiest way for detecting the B19-T500 Trojan is to write some assertions for checking the IR bits directly, this style is not practical for two reasons: First, in a realistic scenario, we do not have any information about the locations of HTs. Even taking this fact into account, thousands (or more) of this type of assertion would be needed for covering all the necessary checks of individual signals, even for small circuits. Furthermore, this style of assertion writing violates the second condition of being a good security checker: it does not describe any notion of a system-level behavior.

Table \ref{tab1} contains the top-level assertions considered for detecting Trojans in the B19-T500 circuit. They have been written in PSL. As shown in the table, the assertions check the correctness of transactions between the memory and processor. The first two assertions check the violation (invalid access) of the write operation in the memory while the next ones do the same check for the read operation. For more details about the mechanism of the memory access in the Viper processor, the reader is referred to its documentation \cite{viper}. Our simulation results show that our assertions can effectively detect the Trojan inserted in the B19-T500 benchmark. The effectiveness of these assertions is discussed later in Section \ref{e}.


\begin{table}[t]

\caption{Considered assertions for detecting Trojans on B19-T500 benchmark }
\vspace*{-4mm}
\begin{center}
\begin{tabular}{@{}lc@{}}
\hline 
\T
 \textbf{Name}&\multicolumn{1}{c}{\textbf{Assertion definition}} \B \\
\hline 
\T
 {{ASR\_1}}&\multicolumn{1}{l}{ \texttt{assert always \{(!(IR == OP\_STORE)) -> (!wr)\}; }} \B\\

 {ASR\_2}&\multicolumn{1}{l}{ \texttt{assert always \{(IR == OP\_STORE)     -> (wr)\};  }} \B\\

{{ASR\_3}}&\multicolumn{1}{l}{ \texttt{assert always \{(!(IR == OP\_READ)) ->  (!rd)\}; }} \B \\

{{ASR\_4}}&\multicolumn{1}{l}{ \texttt{assert always \{(IR == OP\_READ)     -> (rd)\}; }} \B \\
\hline 

\end{tabular}
\label{tab1}
\end{center}
\vspace*{-8mm}
\end{table}

\subsection {Binding the assertions to the main design}
While simulation provides useful information about the incorrect behavior of the circuit and different internal values, it is not sufficient for determining the performance characteristics of the design such as power, timing, and area. Hence, it is impossible to qualify the assertions via simulations only. Instead, the design should be synthesized, and exact performance reports being taken into consideration. As mentioned earlier, PSL and SystemVerilog are the most commonly used languages for describing assertions, but these assertions are not directly synthesizable. Therefore, we have to transform these assertions to a synthesizable format such that the performance results can be obtained after synthesis. For this purpose, we use the MBAC tool \cite{a10} to convert PSL and SystemVerilog assertions to a synthesizable Verilog format. 

After generating a synthesizable code from the assertions, they can be bound to the main circuit so that we can evaluate the effectiveness of the assertions based on the overheads imposed on the circuit. For this purpose, we first synthesize the main circuit without the assertions and obtain the maximum clock frequency, power, and area reports. Later, the original circuit and the bound assertions is once again synthesized. At this point, the assertion goes from a verification asset to an embedded online checker. Finally, these results of the two syntheses are compared to evaluate the performance of the assertions.

\subsection{Security Coverage}

Despite having information about the performance results for each assertion, we cannot still reach a complete decision regarding the effectiveness of a given assertion. In other words, although we know exactly the cost of these assertions, we are not aware of what we achieve. So, a new evaluation scheme is needed to build a trade-off between the costs and the achievements. In this sense, we propose a new metric for the assessment of the assertions considering the security properties. The general idea is to synthesize the original design along with the assertion circuit and check if there are any \textbf{functional paths}\footnote{A functional path is one that can be exercised with a combination of valid inputs. This is in contrast with structural paths or timed paths (STA) that might not be reachable.}  from the individual nodes in the main design to the output of the assertion circuit. Cadence JasperGold Security Path Verification (SPV), our tool of choice, performs proofs to find the existence of functional paths between the design nodes and the assertion outputs, and existence of such paths means that the origin nodes are covered by the assertion. Therefore, if the origin node in the design is the location of the payload of a Trojan, an assertion that can be reached by that node can detect the malicious logic. A higher number of nodes reachable from the original circuit to the assertion output represents better coverage in Trojan detection for that assertion. Based on these explanations, we define our security coverage metric as follows:
\begin{equation}
Security \: Coverage= \frac{ Number \:  of \:  reachable \:  nodes}{Number \:  of \:  total \:  nodes}
\label{eq}
\end{equation}

To obtain the security coverage for each assertion, first, the circuit is synthesized while the assertion is bound to it. Then, all the nodes in the synthesized netlist have to be extracted for further analysis. For this purpose, a tool is developed which receives a netlist as an input and generates a list containing all nodes inside it. This list is then submitted to SPV. This tool is mainly used to check if a part of a design is securely isolated from the other parts, usually referred to as taint analysis (e.g., test if features of a processor should not be accessible except in the test mode), but with some changes in its initial configuration, it can be utilized to calculate the security coverage we need. For this aim, we create a list of pairs of nodes in the format {\emph{origin}, \emph{destination}}, where all nodes in the circuit are possible origin nodes and a destination node is the output of an assertion. Then, the list of pairs of nodes is fed into the SPV tool to check if there is a functional path between them. In this stage, the inner engines of the SPV tool create properties for each pair and try to prove that there is no functional path for the property or to provide some counterexamples for the opposite condition. After finishing the analysis, the security coverage can be calculated using equation \ref{eq}. A Tool Command Language (TCL) script is used to automate the process. 

Now, the needed information for evaluating the effectiveness of the assertions can be taken into account for qualifying them regarding the security aspects. The security coverage can be utilized to drive a trade-off analysis and help the user to decide which assertions are suitable for his/her purposes. It should be noted that not all the circuits need 100\% of security coverage; for example, if some sensitive parts of the circuit are already identified and are the only parts needed to be secured, covering those parts is sufficient to satisfy the user demands.

\section{OpenTitan - a case study}\label{c}
In the previous section, we studied our own defined assertions to prove that they can detect Trojans and be seen as security checkers. But, this practice is hard to generalize: writing such top-level assertions is significantly time-consuming and hard to achieve. Moreover, the main contribution of this work is precisely to reuse the assertions that already exist for verification purposes, instead of generating new ones. Hence, in this section we study the evaluation of different assertions written for verifying the OpenTitan System-on-Chip (SoC) \cite{ot}, to check if they can be used as security checkers.

OpenTitan is an open-source project consisting of a RISC-V-based processor and IPs from different vendors \cite{ot}. It also includes functional assertions for different Intellectual Properties (IPs) which makes it a remarkable candidate for our study\footnote{Some assertions have simulation-based nature and cannot be synthesized (i.e., assertions checking whether a signal is \textit{X} or not).}. 

For this purpose, the Register Top modules of each IP are chosen. This module controls the transactions between the IP and the bus, and is responsible for granting access to read/write requests for IP registers. Moreover, it has a unique error generation mechanism for writes and reads that target addresses that are not represented within the register list \cite{ot}. Since the Register Top modules of different IPs include the same assertions, it provides a good comparison among the experiments. A set of selected assertions is shown in Table \ref{tab4}. In total, 108 different assertions are studied on 35 individual IPs in OpenTitan SoC to demonstrate that the obtained results are comparable to a realistic example and our approach is scalable to industry-size circuits.

To obtain the security coverage of each assertion, the same flow as explained in the previous section is used: MBAC translation, assertion binding, and then synthesis. Finally, the nodes from the synthesized netlist are fed into the SPV tool to calculate the security coverage.

\begin{table*}[t!]
\caption{Considered assertions for Register Top modules of different IPs in OpenTitan SoC }
\begin{center}
\scalebox{0.92}{
\begin{tabular}{@{}l l@{}}
\hline 
\T
 \textbf{Assertion name}&\multicolumn{1}{c}{\textbf{Assertion definition}} \B \\
\hline

\T
 \texttt{\textbf{wePulse}}&{\texttt{assert property (@(posedge clk\_i) disable iff ((!rst\_ni) !== 1'b0) \$rose(reg\_we) |=> !(reg\_we));}}\\

\T
 \texttt{\textbf{rePulse}}&{\texttt{assert property (@(posedge clk\_i) disable iff ((!rst\_ni) !== 1'b0) \$rose(reg\_re) |=> !(reg\_re));}}\\

\T
 \texttt{\textbf{reAfterRv}}&{\texttt{assert property (@(posedge clk\_i) disable iff ((!rst\_ni) !== 1'b0) \$rose(reg\_re || reg\_we) |=> tl\_o);}} \\

\T
\multirow{2}{4em} {\texttt{\textbf{en2addrHit}}} &{\texttt{assert property (@(posedge clk\_i) disable iff ((!rst\_ni) !== 1'b0) ((reg\_we || reg\_re)}} \\
{} & {\texttt{|-> \$onehot0(addr\_hit)));}} \B \\
\hline
\end{tabular}}
\label{tab4}
\end{center}
\vspace*{-8mm}
\end{table*}

\section{Optimizing the assertion list}\label{d}

\begin{figure}[t]
\vspace*{-5mm}
\includegraphics[max size={10 cm}{8.5 cm }, center ]{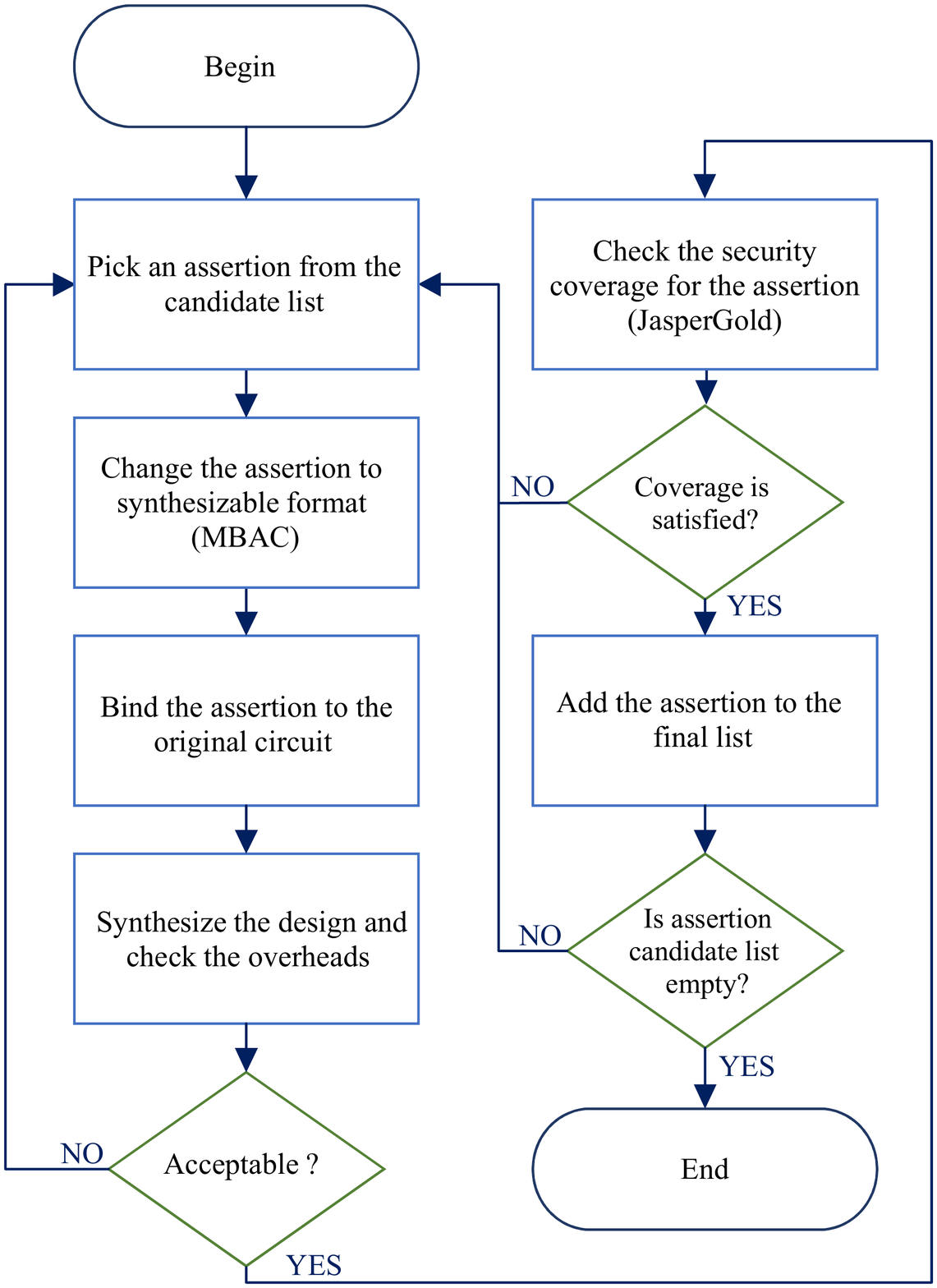}
\vspace*{-12mm}
\caption{Proposed flow for optimizing the assertion list}
\vspace*{-6mm}
\label{fig3}
\end{figure}

Manually checking the assertions to see if they are top-level or not is a very time-consuming process and it questions the efficiency of the proposed approach. Hence, a decision flow is needed to wisely choose the assertions suitable for the security aims.
For this purpose, we present a methodology to help the user only pick efficient security checkers from the available assertions based on his/her needs. This is a necessary step since all the functional assertions are not suitable for security purposes. The overall flow of this methodology is shown in Fig. \ref{fig3}. The first step is to create a list of candidates containing the assertions which can be recognized to be synthesized along with the original circuit, and pick one. Then it has to be converted to a synthesizable format (step 2) to be bound to the main design (step 3). After the synthesis process, different performance reports are generated to help the user decide if the overheads are acceptable or not (step 4). The margin threshold for the overheads can be defined by the user based on his/her needs, and if the overheads for the selected assertion go beyond the defined boundaries, that assertion is put away and another one is picked from the candidate list. Otherwise, the circuit nodes are extracted and fed into the SPV tool to obtain security coverage (step 5). Finally, the user can decide to add this assertion to the security checker list based on the trade-off between the overheads and the security coverage or to select another one from the candidate list.

\subsection{Optimization flow for selecting the assertions}
In this part, we choose Alert Handler IP from the OpenTitan SoC \cite{ot}, which contains several assertions and we show how to form a list of security checkers among these assertions. At the first step, a candidate list including 13 different assertions is created. These assertions are predefined by the OpenTitan developers and their main objective is to make sure that the functionality of the circuit will remain the same as its intent. Since the final security checker list is defined based on the user needs, we defined two different strategies for selecting the appropriate candidates: \textit{fixed-threshold} and \textit{dynamic-threshold} strategies. It should be noted that defining the strategies is completely flexible and depends on the amount of security needed in the cost of performance reduction. In the following, we explain the proposed optimization flow for our defined strategies.

\textit{\textbf{Fixed-threshold strategy:} If the maximum performance overhead for each assertion is X (percent), the security coverage should be at least 10X (percent).}

This strategy defines fixed thresholds for the overheads and/or security coverage of each assertion, and removes the assertions that violate these thresholds from the candidate list. To follow the strategy rules, different overheads of each assertion should be obtained first, and after calculating the security coverage, the maximum overhead (area, power, or timing) goes under comparison. Based on this strategy, only 2 out of 13 assertions of the Alert Handler IP are removed from the candidate list.  


\textit{ \textbf{Dynamic-threshold strategy:} The maximum performance overhead for each assertion should not exceed twice the average performance overheads. For the security coverage, we only pick those assertions that have a positive impact on the overall coverage of the circuit compared to other assertions.}

In contrast to the previous strategy, where assertions are assessed individually, the dynamic strategy performs comparisons between competing assertions. We follow the same procedure to obtain the performance overheads for the first condition of this strategy, but by looking at the security coverage results for individual assertions, no information can be obtained regarding the positive impact of the assertions. Instead, this strategy selects only assertions that fare better than average. 

The first strategy, while simple and easy to implement, requires the user to define a constant (10) for the threshold. The second strategy requires no such constant, but a sufficient number of assertions is needed for defining what average overhead and coverage look like. More details are provided in the next section, where we show how the dynamic strategy can be more effective than its fixed threshold counterpart. Nevertheless, more convoluted strategies can be defined and this remains as future work.

\section{Experimental Results}\label{e}

In this section, we present experimental results including performance overheads and security coverage obtained for different circuits as explained in previous sections. For all experiments reported here, we have used Cadence Genus and our target cell library is a commercial 65nm library. 

\subsection{B19-T500 benchmark from Trust-Hub}
\begin{figure}[t]
\vspace*{-3mm}
\includegraphics[max size={5cm}, center]{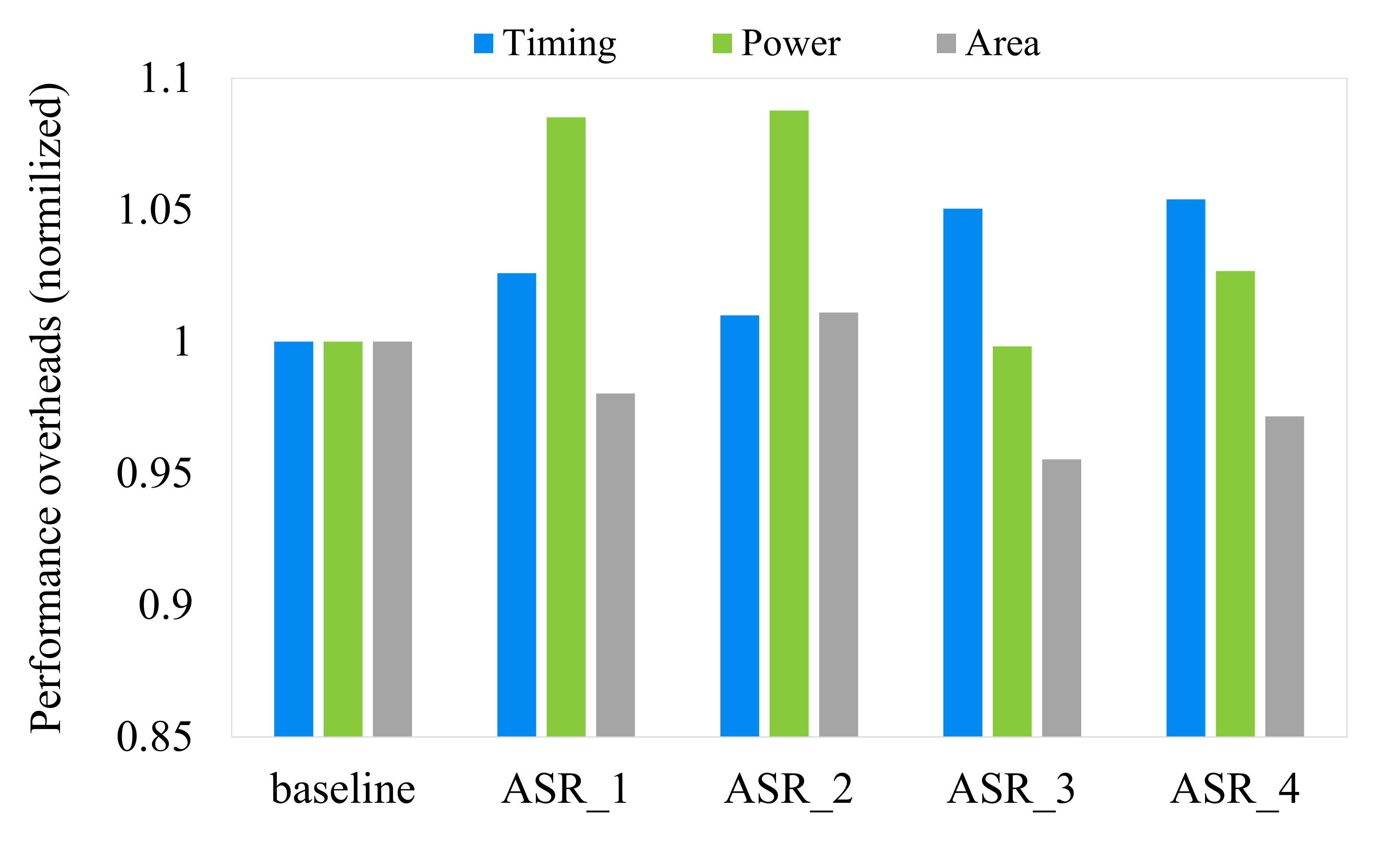}
\vspace*{-9mm}
\caption{Performance overheads imposed by different assertions of B19-T500 benchmark from Trust-Hub}
\vspace*{-6mm}
\label{fig1}
\end{figure}

Fig. \ref{fig1} shows the normalized numbers of timing, power, and area overheads for the assertions considered for the B19-T500 benchmark. As shown in this figure, while the area overhead for three of the assertions (\textit{ASR\_1, ASR\_2,} and \textit{ASR\_4}) is zero, the maximum overheads belong to \textit{ASR\_2} and \textit{ASR\_1} respectively, which consume approximately 9\% more power than the original circuit. Also, the timing overhead is less than 6\% for all of the assertions. It should be noted that the normalized numbers lower than 1 are within the noise margin of the circuit and do not have any effect on the performance.

Table \ref{tab3} shows the security coverage calculated for the same assertions in the B19-T500 benchmark. As shown in this table, our assertions cover on average 6.8\% of the total nodes in the circuit, which means that they can catch the Trojans in their covered areas, regardless of how rare the Trojans are triggered and what impacts they would cause to the circuit. This is one of the main advantages of our method such that the user has no more to be concerned about activating the rare Trojans.

\begin{table}[b!]
\vspace*{-6mm}
\caption{Converted assertions to synthesizable format using MBAC tool}
\vspace*{-4mm}
\begin{center}
\scalebox{0.91}{
\begin{tabular}{@{}l c c c }
\hline 

\rule{0pt}{10pt} \textbf{Assertion name} & \textbf{Total nodes} & \textbf{Covered nodes} & \textbf{Security Coverage} (\%) \B  \\
\hline 

\rule{0pt}{10pt} {ASR\_1}&{ 5014}& { 315} &{6.28}  (\%) \\

\rule{0pt}{10pt} {ASR\_2}&{ 5062}&{ 304} &{6.01} (\%) \\

\rule{0pt}{10pt} {ASR\_3}&{ 4916}&{ 367} &{7.47} (\%) \\

\rule{0pt}{10pt} {ASR\_4}&{ 4944}&{ 369} &{ 7.46} (\%) \B \\
\hline \hline

\rule{0pt}{10pt}  \textbf{Average}& \textbf{ 4984}&  \textbf{ 339} &{  \textbf{6.80}} (\%) \B \\
\hline 

\end{tabular}}
\label{tab3}
\end{center}
\vspace*{-6mm}
\end{table}

\subsection{OpenTitan SoC}
Fig. \ref{fig2} depicts the security coverage obtained for different IP Register Top modules of OpenTitan SoC. As shown in this figure, the highest number for security coverage is 4.77\% for the \textit{nmi\_gen\_reg\_top} module, and the security coverage percentage for the majority of IPs is less than 1, which does not represent a good candidate for being a security checker. This is mainly because these types of assertions are only performing small interface checks, and do not satisfy the condition of describing the top-level behavior of the circuit. Instead, they only cover some local nodes which leads to a low security coverage for the whole circuit nodes. This justifies the need for the optimization step in our proposed methodology to avoid selecting unnecessary assertions that do not have a considerable impact on Trojan detection.

\subsection{Selecting the Assertions}
In this part, we present a practical experiment using the optimization flow as shown in Fig. \ref{fig3}. We defined two strategies for selecting the proper assertions in Section \ref{d}, and in the following, we provide more details about the procedure of assertion selection.

Fixed-threshold strategy: The performance results for the assertion candidate list are shown in Fig. \ref{fig4}. As shown in this figure, the maximum number for the overheads belongs to timing degradation of \textit{ah\_asr\_8} assertion (2.99\%), while the minimum overhead belongs to \textit{ah\_asr\_3} and \textit{ah\_asr\_4} assertions with the number of 0.75\%. At the next step, they should be checked for the security coverage which makes it to be at least 10 times higher than the maximum overhead for each assertion. Fig. \ref{fig5} shows the security coverage results obtained from each assertion using the SPV tool. For better understanding, we simply associated numbers with the assertion names. Based on these results, we can ignore the \textit{ah\_asr\_12,} and \textit{ah\_asr\_13} since they do not satisfy the required security coverage condition and consider the other candidates as the final security checkers. Although 15\% of the assertions were removed based on this strategy, defining smarter strategies can enhance the effectiveness of the final list. Hence, second strategy is defined on the same candidate list for achieving more efficiency.

\begin{figure}[t]
\vspace*{-6mm}
\includegraphics[max size={9.5cm}{10.4cm} ,left]{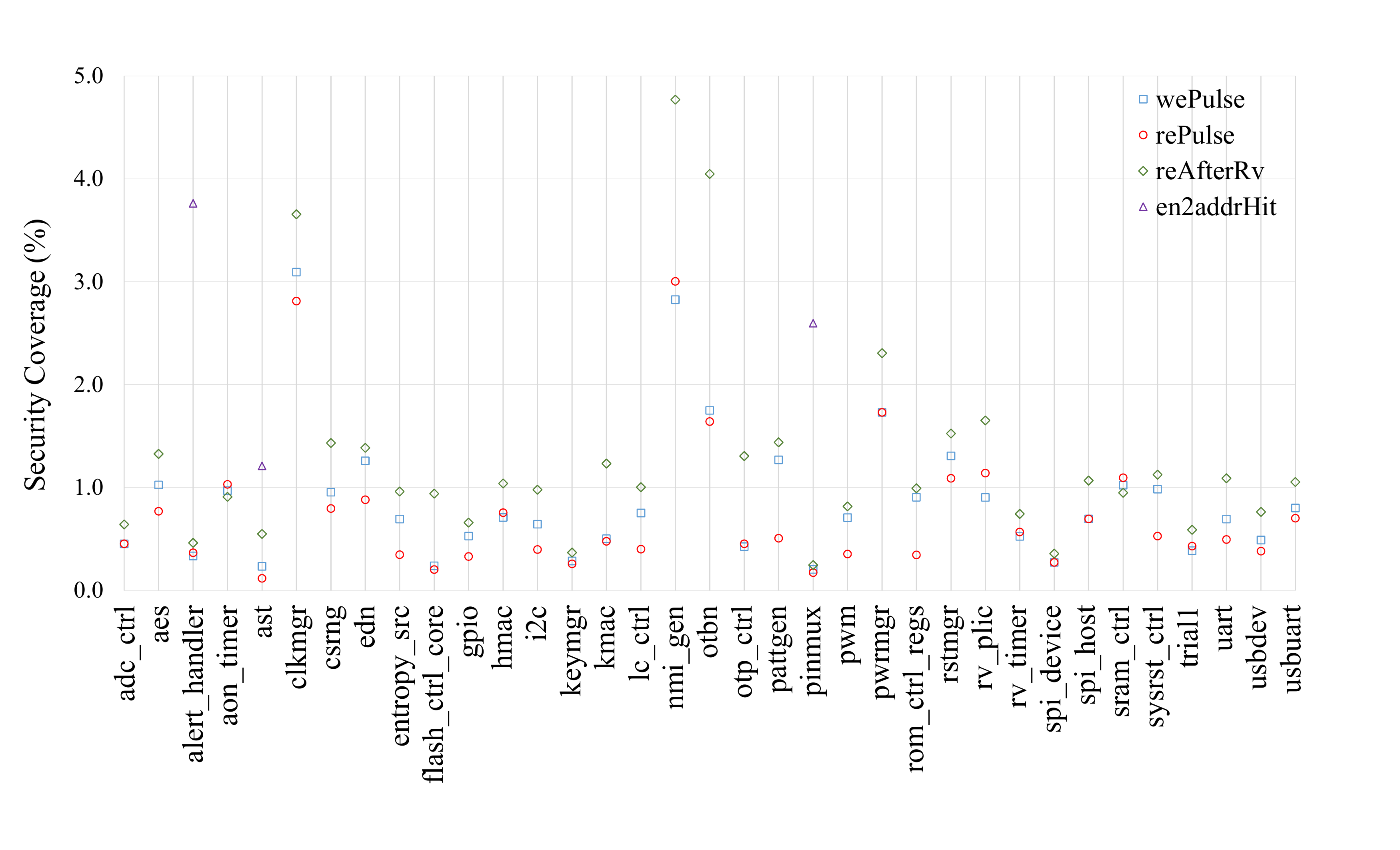}
\vspace*{-12mm}
\caption{Security coverage percentage for the Register Top modules of OpenTitan SoC IPs}
\label{fig2}
\vspace*{-7mm}
\end{figure}

\begin{figure*}[ht!]
\minipage{0.32\textwidth}
  \vspace{-3mm}
  \includegraphics[width=\linewidth]{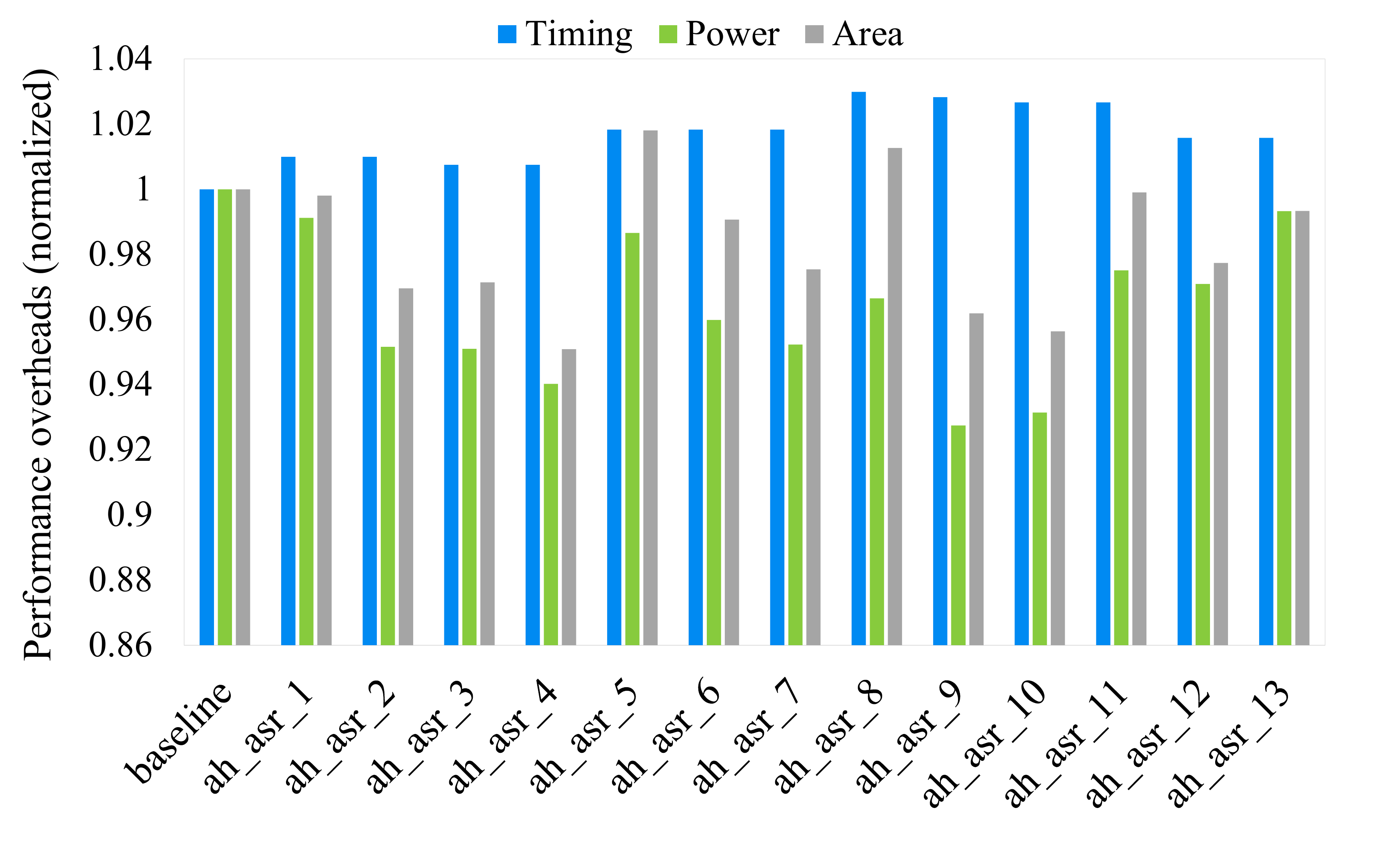}
  \vspace{-9mm}
  \caption{Performance overheads imposed by different assertions of Alert Handler IP}\label{fig4}
\endminipage\hfill
\minipage{0.32\textwidth}
  \vspace{-2mm}
  \includegraphics[width=\linewidth]{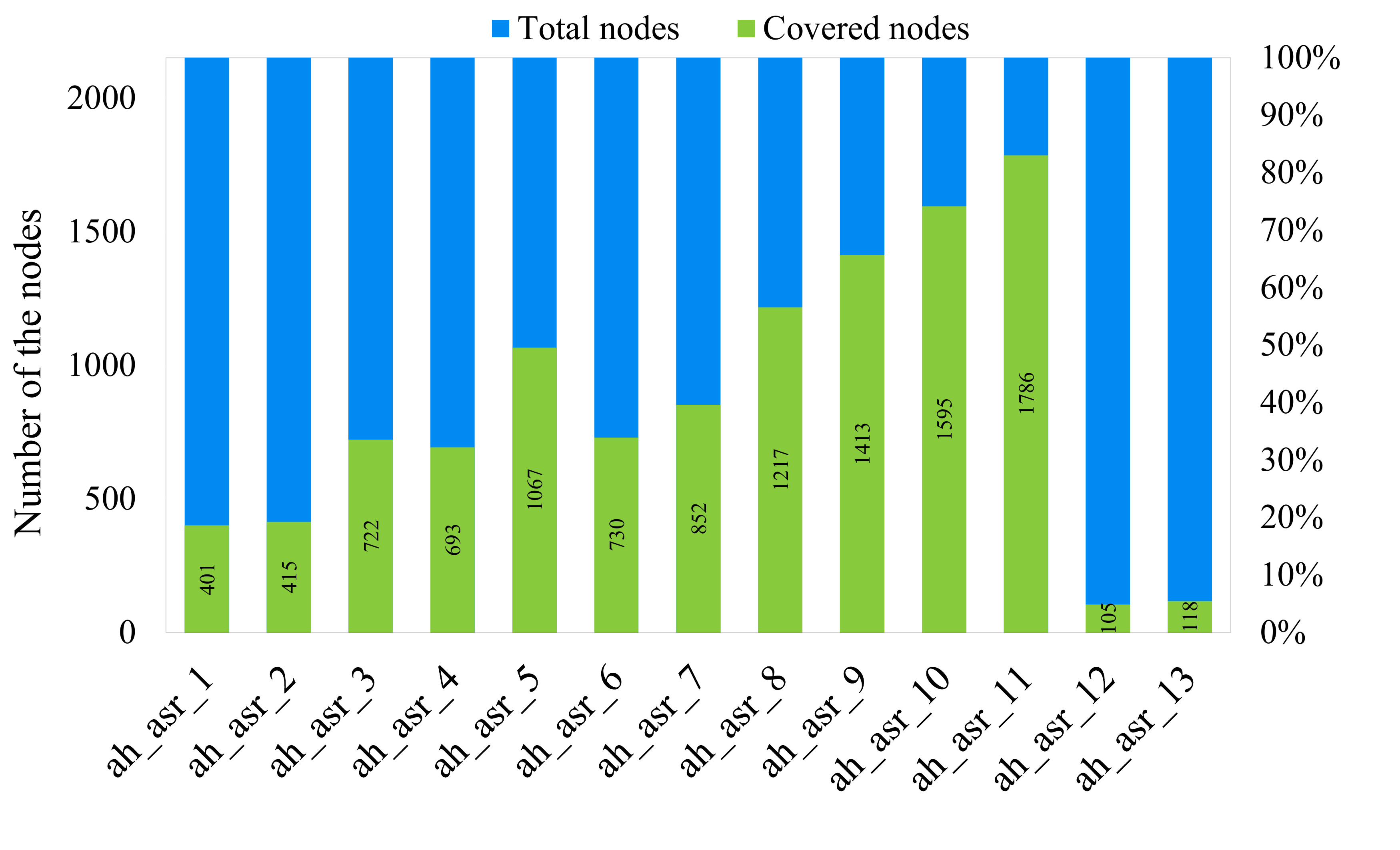}
    \vspace{-10mm}
  \caption{Number of covered nodes for the individual assertions of Alert Handler IP}\label{fig5}
\endminipage\hfill
\minipage{0.32\textwidth}%
  \vspace{-3mm}
  \includegraphics[width=\linewidth]{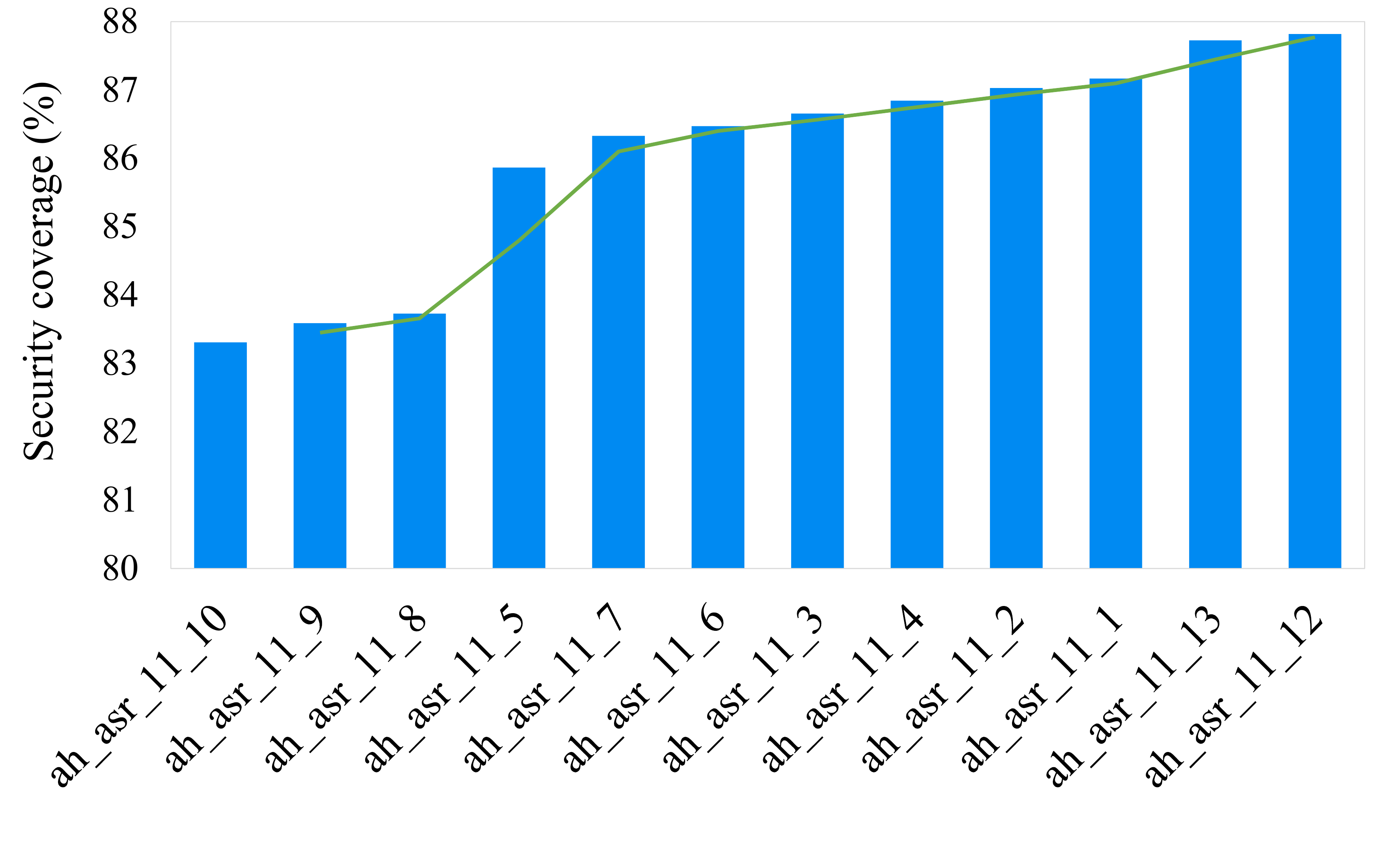}
    \vspace{-10mm}
  \caption{Security coverage percentage for different assertion sets of Alert Handler IP}\label{fig6}
\endminipage
 \vspace{-6mm}
\end{figure*}

Dynamic-threshold strategy: For the first condition of this strategy, we should calculate the average overhead for all the assertions which is 1.79\%. Hence, all the candidates are passed since they have less than twice the average overhead in all the cases (Fig. \ref{fig4}). But for the second condition, it is not sufficient to refer to the security coverage results since they do not have any notion of comparison to each other. Hence, an extra step is required to select the security checkers. For this purpose, we organize the assertions from the highest security coverage (\textit{ah\_asr\_11})  to the lowest one (\textit{ah\_asr\_12}). Starting from the \textit{ah\_asr\_11}, we add the assertion with the next highest number (\textit{ah\_asr\_10} in the first round) and calculate the security coverage for the new set of assertions we made. Then, the next highest number is added to the existing set, and this process is repeated until the lowest number is added to the list. 

Fig. \ref{fig6} represents the security coverage numbers for each set of assertions. We include the assertion numbers in naming the set of assertions to identify the effect of added assertion in each step. For example, \textit{ah\_asr\_11\_10} represents a set of assertions that starts from \textit{ah\_asr\_11} (the highest coverage) and ends with \textit{ah\_asr\_10} (the last assertion added to the list), and \textit{ah\_asr\_11\_9} includes the assertions \textit{ah\_asr\_11}, \textit{ah\_asr\_10}, and \textit{ah\_asr\_9}. Furthermore, a moving average trend-line is added to this figure to help for choosing the best assertions. Since the period of the moving average trend-line is 2, it can make a good comparison between the security coverage of the newly added assertion in each stage, and the number for the two previous assertions. Therefore, if the security coverage obtained after adding an assertion crosses the moving average trend, it can be realized that a noticeable difference has happened. Returning to the second condition of Strategy 2 and from Fig. \ref{fig6}, we can see that security coverage numbers of only three assertions have crossed the moving average trend-line (\textit{ah\_asr\_5, ah\_asr\_7,} and \textit{ah\_asr\_13}). Hence, they can be added to the final list. Moreover, the \textit{ah\_asr\_11} assertion is added to the final list since it has the highest security coverage. 

In contrast with the results of the Register Top modules of different IPs (Fig. \ref{fig2}), the security coverage numbers of different assertions in Alert Handler IP are relatively higher (Fig. \ref{fig5}). This is mainly because the assertions written for this specific IP are describing a top-level behavior of the design, rather than checking only local signals and interfaces. 

As shown in these two examples, different strategies can be defined based on user needs which makes the presented approach flexible. Moreover, one of the advantages of our work comparing with the current approaches is the simplicity of using it without complex procedures. For example, the presented work in \cite{t} supports Trojan detection with flexible overheads, but it requires a lot of effort and complicated steps. In contrast, we use commercial tools that are available to the community, thus increasing the portability and scalability of the presented work.

\section{Conclusion}\label{f}
In this paper, we presented a new methodology for using verification assertions as security checkers. The security coverage, our proposed metric for assessing the effectiveness of assertions in Trojan detection, abstracts the notion of a Trojan trigger and focuses on the effect of the payload. 

We examined our methodology on case studies from the Trust-Hub benchmarks and the OpenTitan SoC with more than 100 assertions to show the scalability of our work for the industry-size circuits. Moreover, we showed how defining a smart strategy can enhance the assertion selection process. In the future, we will focus on automating these strategies to enhance the current methodology.

\section*{Acknowledgment}

This work was partially supported by the EU through the European Social Fund in the context of the project “ICT programme". It was also partially supported by the Estonian Research Council grant ``MOBERC35".

\bibliographystyle{IEEEtran}
\bibliography{ref}

\end{document}